# Defining UML Family Members Using Prefaces

Steve Cook
IBM European Object Technology Practice, UK
sj_cook@uk.ibm.com

Anneke Kleppe
Klasse Objecten, The Netherlands
a.kleppe@klasse.nl

Richard Mitchell
University of Brighton, UK
richard@inferdata.com

Bernhard Rumpe
Technische Universität München, Germany
rumpe@in.tum.de

Jos Warmer
Klasse Objecten, The Netherlands
J.Warmer@klasse.nl

Alan Cameron Wills
TriReme International, UK
alan@trireme.com

*Abstract*

*The Unified Modeling Language is extensible, and so can be regarded as a family of languages. Implicitly or explicitly, any particular UML model should be accompanied by a definition of the particular UML family member used for the model. The definition should cover syntactic and semantic issues. This paper proposes a mechanism for associating models with such definitions. Any particular definition would form what we call a preface. The name is intended to suggest that the definition of a particular UML family member must conceptually come before any model built using that family member. A preface would be large, and should be organised using packages. This would allow large amounts of sharing between different prefaces. The paper proposes that prefaces should have an axiomatic style of semantics, though not necessarily fully formal, and it offers a general approach to semantics that would reduce problems of inconsistency within a large preface, based on the idea of general cases and special cases.*

# 1. Introduction

The Unified Modeling Language (UML) is a standard language for modelling object oriented software systems (Booch, Rumbaugh and Jacobson 1998). The UML consists of a variety of notations to describe different views of a system, at different levels of abstraction. The precise syntax and an informal definition of the meaning of the UML notations are given in (OMG 1997), where they are presented in a mixture of diagrammatic and textual languages, including the UML's own Object Constraint Language, OCL (Warmer and Kleppe 1999). The semantics are neither complete nor fully precise.

By design, the UML is an extensible language. So the UML can actually be regarded as a family of languages, embracing a variety of syntactical specialisations and extensions through stereotypes and tags, as well as adopted meanings suited to different application areas and domains. Modellers using the UML are therefore free to make certain choices about syntax and semantics. This makes it difficult to define a precise semantics of UML for all cases, and places a responsibility on modellers to tell those who read their models which choices of syntax and semantics were made. This gives the modeller the freedom and also the responsibility to tailor the UML to his or her specific needs. At present, the UML does not offer a suitable mechanism to allow modellers to describe these choices. This paper proposes such a mechanism, using what we call prefaces.

A preface defines the syntax and semantics of the modelling language used to build any particular model. In other words, a preface defines one member in a family of modelling languages. The name 'preface' was chosen to suggest that the definition of a modelling language should be thought of as coming in front of any particular model written in that language.

A preface would typically be a large document. It could be organised into packages, and use the UML's package import mechanism. Importing existing preface packages would allow a modeller to choose a certain UML family member, containing particular predefined stereotypes, restrictions on the use of some notations, and so on, and giving a specialised meaning to modelling concepts. Writing new packages would allow a modeller (or, more likely, a method designer or a tool builder) to define a new member of the UML family, building on existing work. Prefaces would be especially powerful if tools provided specific assistance for them.

Prefaces would allow agreement to be reached and documented—a group of modellers could agree to adopt the same set of preface packages. And prefaces would support variation—an individual tool vendor could define some special packages to accompany a tool designed to meet his or her clients' particular needs.

Prefaces could be developed from current definitions of the UML. And they could build on existing work on semantics such as (France, Evans, Lano and Rumpe 1998), and (Breu, Hinkel, Hofmann, Klein, Paech, Rumpe and Thurner 1997). This latter work discusses the problems of fixing a single semantics for the UML.

The proposals contained in this paper reflect the authors' experience in developing, using, analysing, defining, and teaching object-oriented modelling notations, techniques and methods, as well as insights gained through their participation in the UML standardisation process.

The remainder of this paper is organised as follows. Section 2 explains more fully why the UML is a family of languages, and argues briefly that this is desirable. Section 3 presents an example of what a preface to a UML model might address. The example concerns the relationship between state diagrams and class diagrams. Then section 4 explores the range of issues that a preface could contain. Prefaces will be large, and the need to organise them is addressed in section 5. Sections 6 and 7 are concerned with semantic issues. Section 6 advocates an axiomatic style of semantics. Section 7 proposes an approach to dealing with inconsistencies between different packages within one preface. A useful extension of prefaces so that a model's meaning would depend on non-functional requirements is outlined in Section 8. Section 9 summarises the paper.

## 2. The UML: a family of languages

Many of those who work in the IT industry build models. Often, these models must be communicated to other people, perhaps across time and space. The builder of a model can communicate that model more clearly to a reader of the model if they have a common understanding of the modelling language they are using—both its syntax, and the semantics of the syntactical elements.

Rather than defining a modelling language each time they need to communicate, a model-builder and a model-reader could agree to use a standard modelling language, with a single, defined meaning. As the UML effort has shown, it is impossible to obtain world-wide agreement on what elements a single object-oriented modelling language should contain, and what the elements all mean. As a consequence, the UML has been designed to be extensible, through such mechanisms as stereotypes and tagged values. As UML's lead designers say, "You can totally change the basic notation of the UML with stereotypes" (Booch, Rumbaugh and Jacobson 1998, pg. 89). Thus, even if model-builder and model-reader restrict themselves to the UML, they still have to agree on which extensions and interpretations of the UML notations they are using. So, in practice, the UML is actually a variety of notations, with a variety of meanings, and with a variety of tools implementing those meanings.

In our judgement, it would be a mistake to try to remove all the variety. A number of groups have good reasons for wanting it. For example, modellers want some variety because different kinds of software need different kinds of models. Attribute grammars are useful for compiler writers, spatial primitives are useful for designers of geographical information systems, and timing diagrams are useful for control engineers. Tool builders want variety because they can differentiate their products by giving them particular capabilities, based on special language features. For example, generating code from class diagrams for a specific framework, such as IBM's San Francisco (IBM 1999), a particular environment, such as Sun's Jini (Sun 1999), or a given database, allows the tool vendor to include certain facilities, like persistency, a transaction mechanism, error recovery techniques, and default implementations of certain interfaces needed by the environment. This considerably specialises the intended interpretation of the diagrams. And it probably needs additional syntax, in the form of tagged values and stereotypes to control the code generation.

In parallel with the work described here, the Object Management Group has explicitly recognised the need for a family of UMLs by its recent (at the time of writing) issuing of Requests for Proposals asking for UML Profiles for CORBA, for Enterprise Distributed Object Computing, and for Scheduling, Performance and Time. Within each Request for Proposal (RFP) there is a working definition of a UML profile, which states: "For the purposes of this RFP, a UML profile is a specification that does one or more of the following:
- Identifies a subset of the UML metamodel (which may be the entire UML metamodel)
- Specifies "well-formedness rules" beyond those specified by the identified subset of the UML metamodel. "Well-formedness rule" is a term used in the normative UML metamodel specification to describe a set of constraints written in UML's Object Constraint Language (OCL) that contributes to the definition of a metamodel element.
- Specifies "standard elements" beyond those specified by the identified subset of the UML metamodel. "Standard element" is a term used in the UML metamodel specification to describe a standard instance of a UML stereotype, tagged value or constraint.
- Specifies semantics, expressed in natural language, beyond those specified by the identified subset of the UML metamodel.
- Specifies common model elements, expressed in terms of the profile."

The proposals for UML prefaces in this paper go significantly beyond this definition of a UML profile in terms of what a preface might cover. At the time of writing, a white paper describing profiles is available to OMG members. Clearly the intent of the OMG's activity on profiles is very similar to the intent of this paper, and over time it may prove appropriate to unify the concepts of *profile* and *preface*. For now, to avoid confusion, we will keep them separate.

Thus, the current situation is that the UML is a family of languages. The likely future is that there will be more variation, not less, as more people use the UML, and as more UML tools emerge. It is therefore crucial to support this variation in the UML family of languages, which, in turn, involves supporting the definition of individual members of the UML family. Furthermore, the definition of a UML family member must be independent of any particular model, so that it can be reused for many models.

An individual family member will not be defined in isolation. The definition of a new family member will usually be expressed as a difference from the definition of one or more existing members. As the UML community's experience with definitions grows, the range of issues addressed in definitions could be expected to grow, and the definition of any particular member might be improved. Therefore, we need to support variation along several dimensions.

## 3. An example of what a preface defines

This section presents a small example of the kind of information that a preface could address. The example is about connecting a statechart diagram to a class diagram (see Figures 1a and 1b). The point of this example is not to evaluate alternative approaches to combining class diagrams and statechart diagrams. The point is that different groups of developers will want to use different approaches, and we are advocating that the diagrams in Figures 1a and 1b should be accompanied by a preface, where those who read the diagrams can find out which approach is being used, and on the basis of which tools can manipulate the diagrams.

Here is one (incomplete) interpretation of the two diagrams. "Because of the numbering of the figures, the statechart diagram is attached to the class diagram, and therefore describes the states of objects of class C. The events on the statechart diagram correspond, by name-matching, to the methods in the class diagram. The statechart diagram shows that objects of class C can be in one of three states, s1, s2 and s3. If an object of class C is in state s1, calling method m1() will cause it to make a transition to state s2, when calling method m2() will cause it to make a transition back to state s1. When it is in state s1 or s2, calling method m3() will put the object into state s3."

In its current form, the UML does not insist on this interpretation. A reader from a real-time background might read the statechart diagram as saying that class C must have an event controller. And someone trained in Syntropy (Cook and Daniels 1994) might read the two diagrams as saying there must be dynamic subtypes of class C for the states s1, s2, s3. So, the model builder must provide additional documentation if he or she wants the diagrams to be interpreted in a particular way. This kind of documentation is what we propose should be written in a preface to any model.

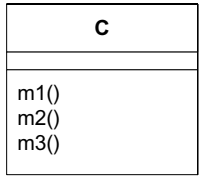
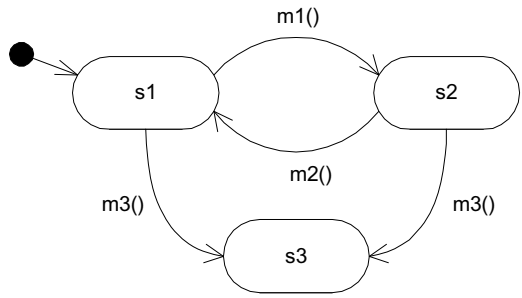

**Figure 1a**. *A class diagram*.  **Figure 1b**. *A statechart diagram*.

We might go further and include in the preface rules for transforming a model in preparation for code generation. The rules could include transferring information from the statechart diagram to the class diagram like this, for example:

1. Each state on a statechart diagram attached to a class induces a Boolean attribute in the class, with the same name as the state.
2. An invariant is induced in the class to constrain these Boolean attributes so that they correctly represent mutually exclusive states.
3. Each event on the statechart is mapped to an operation with the same name in the class.
4. For each transition that is labelled with an event, a precondition is induced in the class to constrain the corresponding operation to start in the state from which the transition originates. If the event originates from several states, the precondition is formed using the logical *or* operator.

A tool that implements these rules could generate Figure 2. It could also generate a skeleton version of the executable code for class C, and even some testing and monitoring code. Of course, this is just an illustrative example and the rules do not form a complete, workable set (for example, the rules do not mention postconditions).

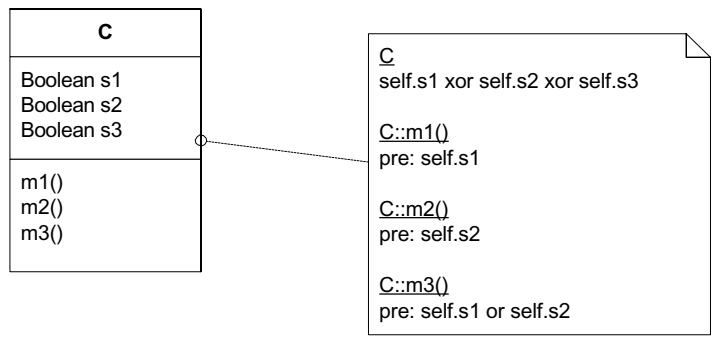

**Figure 2**. *Class C with information induced by the statechart diagram.*

## 4. The content of a preface

This section looks more closely at what a preface might contain. It does not attempt to be comprehensive (indeed, if the idea of prefaces is adopted, there will never be a definitive list of the possible contents of a preface). Instead, it tries to show the enormous range of issues that prefaces might address for the kinds of UML models being built today. The question of how to organise such a wealth of information is deferred to Section 5.

A preface will contain definitions of:
- metalevel concepts
- relations between metalevel concepts
- features of metalevel concepts
- well-formedness rules (i.e., constraints on metalevel concepts).

Thus, a preface can specialise, tailor, and extend the following:
- the metamodel
- the abstract syntax
- the semantics
- the concrete syntax
- the context conditions on all of the above
- the allowed transformations and generations.

The items in this list are expressed using the definite article "the" as a reminder that a given model would have a single preface (describing, for instance, *the* concrete syntax used in this model).

At a more detailed level, here are some of the technical issues that a preface might cover. These are just examples, to give an idea of the range of issues a full definition of a modelling language might address:

- extensions of the UML by new elements through
  - stereotype definitions
  - allowable tags
  - pragmas (expressed as tagged values)
- class diagrams
  - meaning of aggregation: strong or weak?
  - special symbols for better visual perception
  - treatment of n-ary associations (n>2)
  - implicit adding of methods for persistence, Jini-Interfaces, etc.
- statechart diagrams
  - meaning of guards
  - what to attach statecharts to (methods or classes?)
  - what happens if a message cannot be processed? Throw an error? Ignore?
  - relationship of guards and control states to preconditions
  - inheritance of statecharts between superclasses and subclasses
- inheritance
  - allowed forms (multiple, public and private, repeated)
  - meaning of repeated inheritance (C++ rules? Eiffel rules?)
  - redefinition rules for overriding (extending types of parameters, how to deal with pre- and postconditions)

- framing rules to specify which actions do or don't affect which parts of a system, on the
  - specification level
  - operational level
- underlying model of time
  - what is an object? what is a system state?
  - causes of transitions: events, operations, conditions that become valid, timeouts, etc
  - concurrency issues: deadlock, transaction privacy, etc.
- connecting OCL constraints to diagrams
  - how UML-diagram elements become available to OCL expressions
- programming language specific support
  - C++ (with multiple inheritance)
  - Java (without multiple inheritance, but a useful exception mechanism)
  - Smalltalk (with possibility to dynamically change a class during runtime)
  - Eiffel (with support for runtime checking of assertions)
- communication paradigms
  - buffered (asynchronous) communication vs. synchronous vs. procedure call
  - synchronous with deadlocks, or loss of messages not processed in time
  - event notification mechanisms, priority, transmission privacy (encryption) through protocols
- persistence mechanisms
  - automatic persistence for all or some objects
  - predefined mapping to a given data-base
  - logging of object-changes; undo policy
  - predefined transaction-mechanism
  - error recovery

Although there is no sharp distinction, some of the issues are more relevant to people building and reading a model, whilst others are more relevant to tool builders (and, thereby, to the people who use the tools). For example, consider a framing rule that says that, unless an operation specification positively defines that a property changes, it does not change. Most of today's tools would not be able to use such information, but people developing software from a model prefaced with this framing rule could incorporate the constraint into their designs, both implicitly, by writing the code correctly, and, perhaps, explicitly, in the form of software contracts. In contrast, a stereotype to mark certain classes of object as persistent could be used directly by a code generator (i.e., a tool) to relieve designers of the task of writing code to make objects of the class persistent.

## 5. Structuring a preface using packages

Because the range of subjects in a preface is enormous, we must expect that prefaces will be very large. The current definition of the UML (OMG 1997) gives an indication of just how large. Therefore, we propose that definitions be organised into a number of packages. Many of these packages could be world-wide or company-wide standards. You would not need to learn a new preface for every UML model you read.

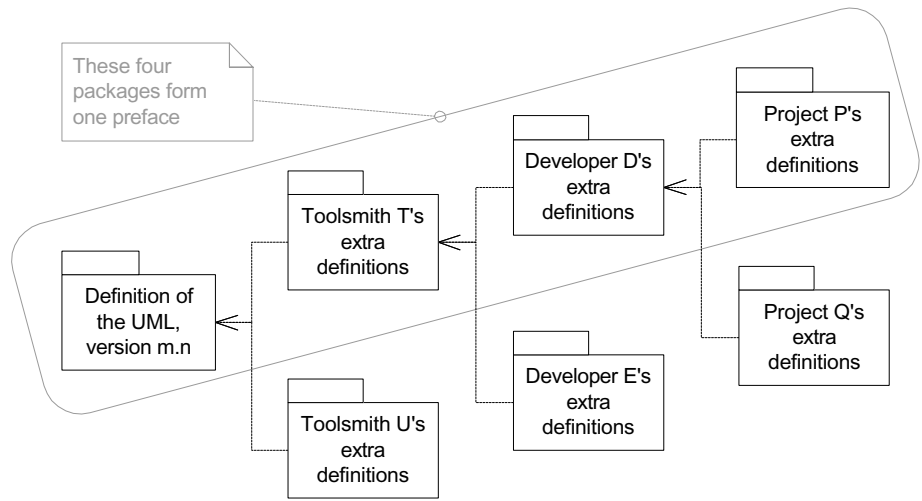

**Figure 3.** *Packages of definitions*

Figure 3 shows a simple example of a hierarchy of packages of definitions (in practice, we would expect the hierarchies to be very much larger). In this example, toolsmith T imports the definitions of some version of the UML, and adds further definitions. Company D imports the augmented set of definitions and adds some company-specific definitions. Finally, project P adds some project-specific definitions. The set of definitions in these four packages form a preface, which would (conceptually) come in front of the model built by the members of project P.

When the model goes through a transition from analysis to design, additional packages could be imported into the model's preface, for example, to define which framework is used as the underlying basis of the implementation. Thus, some important decisions to be made during the development process could be made and documented through selecting appropriate packages and importing them into the preface to the model.

## 6. The semantics of a preface

When you read a model, you draw inferences from it, or try to. You try to work out what is true (and what is not true) from the model in front of you. For example, when you read the model in Figure 1b, you notice that state s1 has arcs leaving it labelled m1() and m3(). You might want to know whether this means that calling m2() on an object of class C that is in state s1 is illegal, or that such a call is legal but has no state-changing consequences, or that the model is unfinished and does not yet specify what calling m2() in state s1 means.

We suggest that the most important goal of a preface should be to tell you how to read a model by defining what you can and cannot infer from its texts and diagrams.

Defining semantics from the point of view of inferences involves defining axioms (things that are given as true statements) and inference rules (rules for inferring more true statements from the ones you already know). This approach is appropriate both for people and for tools. For example, generating a class diagram such as the one in Figure 2 from the diagrams in Figures 1a and 1b can be seen as a form of inference (Evans, France, Lano and Rumpe 1998). Information given in Figure 2, such as the preconditions on the methods, is inferred from the diagrams in Figures 1a and 1b.

Even with a chosen style of definition, there is a question of what form the definition should take. In particular, there is the question of how formal it should be. Different packages within one preface might have different forms, taking into account users' preferences and the kind of tool support available.

In general terms, the axioms and inference rules will cover diagrams, informal explanations, semi-formal elements (such as stereotype names), and more formal elements, such as statements in OCL. In more detail, a preface could be defined using one or more of the following approaches:

- Informally, using natural language and diagrams
    - an informal approach is appropriate for introductions to the UML
    - current UML textbooks can be regarded as examples
    - the recent book on Catalysis (D'Souza and Wills 1999) could be regarded as a package that extends the introductory textbooks, yielding a richer preface
    - more specialised prefaces can be also defined informally (for example, informally describing the meaning of stereotypes introduced for a particular project).
- Rigorously, making some use of formal languages, such as OCL, on the meta-level
    - this approach is appropriate for definitions of UML extensions and specialisations
    - the current UML semantics guide is an example
    - another example would be a tool vendor's definition of what meanings the tool gives to UML models, in terms of what the code generator outputs
- Formally, making full use of formal languages, such as OCL, on the meta-level
    - this approach is appropriate, for example, for computer scientists exploring the foundations of UML
    - a fully formal approach would have a limited audience
    - a fully formal approach would help those building and using tools that support inference and proof
- Implicitly executable, by being hard-coded within a tool
    - the preface is in the form of the tool itself and its user guide
    - current UML tools that generate code are examples
- Explicitly executable, using some kind of scripting language that allows tool users to add functionality
    - this approach is appropriate for tool builders and users
    - scripts can
        -- check certain context conditions in addition to the implemented ones, that ensure certain specific needs of the modeller
        -- infer information about the UML model, such as that needed to build a dictionary
        -- enrich one part of a model with information from another (such as transferring information from statecharts to class diagrams)
        -- guide the generation of code

We are confident that, using prefaces, UML models can be defined more precisely than now without necessarily giving a fully formal semantics to UML family members.

## 7. Overlap and Inconsistency

We have suggested that a preface should be structured into packages. One preface could include standard packages from consortia such as the OMG, tool-specific packages from a tool-builder, company-standard packages, and project-specific packages. As a result, a single preface will usually have packages that come from many sources.

As soon as we allow multiple sources, we must decide what to do if two packages overlap. When two packages overlap, they define the same thing twice. The two definitions might agree, in which case there is no problem. Or they might conflict, and some mechanism for conflict resolution must be applied.

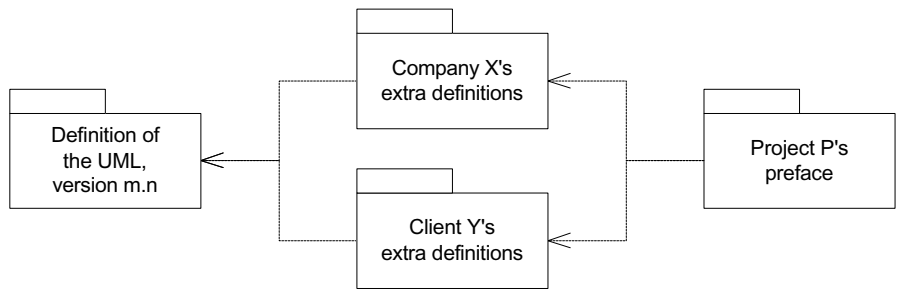

**Figure 4**. *A preface can have potentially conflicting definitions*

A simple solution is to insist that packages do not contain conflicting, or inconsistent, definitions. In practice, this would be unworkable. Packages will be defined in many places and at many times. Imagine you are working on Project P and you are faced with the situation shown in Figure 4. Project P's preface imports some basic UML definitions, and two sets of additions, one from within the company, and one from a client. There might be places where the two sets of definitions conflict. Saying, "well, we won't use one of the conflicting sets" might not be an option. The client might insist that the project uses a particular set of extensions to the UML, and the project team might need to use a company-wide set of extensions to the UML embodied in a tool.

We propose two mechanisms intended to help with the problem of conflicting definitions. First, we propose that a preface is built from an ordered list of packages, so that the following two definitions of project P's preface are not equivalent (the examples use no particular syntax).

```
package Preface-Project-P-version-1
    import
        package "Toolsmith T's extra definitions"
        package "Client C's extra definitions"
    define
        -- body of this package
end

package Preface-Project-P-version-2
    import
        package "Client C's extra definitions"
        package "Toolsmith T's extra definitions"
    define
        -- body of this package
end
```

The order in which packages are imported matters: later definitions are taken as redefinitions of earlier ones. The definitions in the body of a package are taken as coming after all the imported packages' definitions.

Secondly, we propose a model for redefinition that is intended to make it usable. We shall illustrate the approach with two examples.

In the first example, one package defines an integer constant *max* to be 10, and a later package defines the same constant to be 8. (As a matter of good documentation, packages could explicitly list which imported definitions they override.) Because there is an order to the definitions, the final definition of *max* is that 'max=8.' Of course, the person responsible for defining the version of the UML in question needs to be aware of any bad consequences of redefining *max* to be 8.

In the second example, one package defines that 'all objects are persistent.' A later package defines that 'event objects are non-persistent.' These conflicting definitions are resolved by regarding earlier definitions as general definitions and later definitions as more specific definitions, which override earlier ones at their points of conflict. Thus, the two definitions are to be read as saying this:

```
if
    x is an event            -- is x covered by a specific case?
then
    x is not persistent      -- yes - use the specific definition
else
    x is persistent          -- no - use the original, general definition
```

Any package used in any preface can be thought of as being qualified by an imaginary statement saying, "unless a later package says something to the contrary then I define that...."

It is impossible to devise an approach to specifying individual languages so that only "good" or "useful" things can be expressed in the languages. We know that the approach we are proposing can be misused, with awful consequences. However, we do believe that the proposed approach to giving meanings to models can work, and that an imperfect solution can be better than no solution at all. In addition, we assume that the job of combining packages of definitions to build prefaces will not be part of the daily work of modellers, but will be carried out by specialists. Further, we expect that there will be public scrutiny of prefaces, just as there is now public scrutiny of the current definition of the UML, such as that by Schürr and Winter (1997) concerning flaws in the packaging concept, and by Simons (1999) concerning interpretations of use cases.

## 8. A possible extension

In simple terms, the overall information path from a UML model to an executable system is shown by the middle row of boxes and arrows in Figure 5.

On the left of the figure is a UML model, built using the syntactic elements defined in the preface to the model. The preface also gives the modelling language a semantics, and hence gives the model an intended interpretation, or meaning. (The current OMG documents describing the UML are an example of a preface, because they define the syntactic elements of the language, and their intended interpretations, albeit not fully.) Code generation transforms this model-with-meaning into (part of) an executable system.

The output of the transformation cannot be all of the system because UML is not, at the time of writing, an executable notation. But the code generator can be given additional information, such as the coded bodies of methods (shown in the top box in Figure 5).

So far, code generation involves producing an executable system from a UML model, which has a meaning given by a preface, together with some non-UML information such as code bodies.

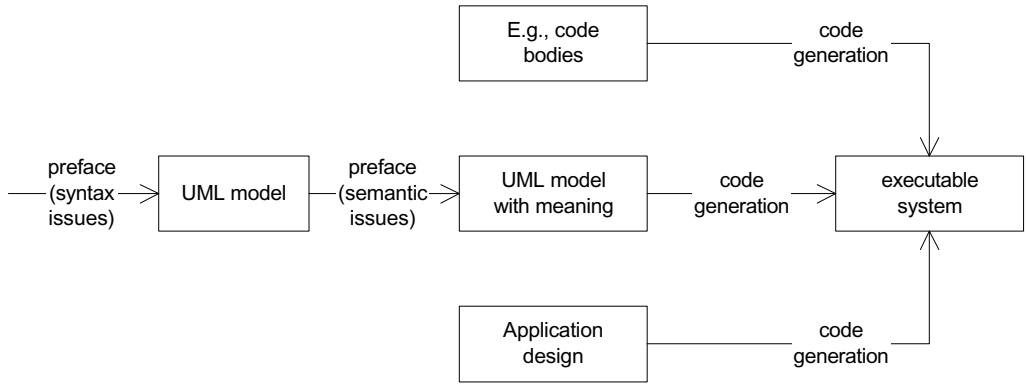

**Figure 5.** *The information path from model to executable system.*

In general, the generated code will also depend on the chosen architecture of the system, and on non-functional requirements such as performance or distribution. For example, the detailed mapping of classes and associations in a UML model onto relational database tables and queries would depend upon such non-functional factors. In Figure 5, we label such factors with the general heading of 'application design.' Our current position is that the preface concept is limited to specifying the semantics of the UML family member in terms of purely functional matters. However, this code-generation mapping has been identified as a focal point for architectural design and several more-or-less formal techniques for describing this mapping have been developed, such as the Hypergenericity concept (Desfray 1994) and Recursive Design (Shlaer and Mellor 1997). We believe it would be fruitful to extend the preface notion in this direction.

## 9. Summary

To summarise, the idea for a preface concept arose from observing that the UML is extensible, and open to different interpretations, and different modellers use it in different ways. So far, there is no agreed mechanism for modellers to tell their readers how to interpret their models.

Prefaces let tool vendors and modellers describe their particular member of the UML family. Prefaces encourage all those who define aspects of the UML to document their definitions. The structuring of prefaces into packages provides a way to combine different people's definitions by importing and extending existing prefaces.

A standard UML preface could contain definitions of the syntax and semantics of a standard core of the UML. Additional preface packages could define variations on this core, allowing extension, adaptation, and tailoring of the UML to be captured. Orienting the semantic style towards inference is appropriate for people, and for tools (which need such things as rules for transforming UML models).

Combining preface packages using the UML import mechanism and allowing certain kinds of overriding provides a general technique to tailor the UML to the specific needs of companies, projects or even individual modellers.

We expect that a very common way for a development team to define their chosen member of the UML family of languages is to state that they are using an existing combination of packages of definitions.

## Acknowledgements

Bernhard Rumpe gratefully acknowledges support from the Bayerische Forschungsstiftung under the FORSOFT research consortium and the DFG under the Leibnizpreis program. Richard Mitchell gratefully acknowledges support from InferData Corporation, and the Distributed Systems and Software Engineering Unit at Monash University.